\documentclass[aps,pre,twocolumn,superscriptaddress,showkeys,longbibliography,floatfix]{revtex4-1}
\usepackage{amsmath,amssymb,amsfonts,amsthm,enumerate,multirow}
\usepackage{color}
\usepackage{siunitx}
\usepackage{hyperref}
\usepackage{natbib}
\usepackage{subfig}
\usepackage{graphicx}
\usepackage{placeins}
\usepackage{dblfloatfix}
\usepackage{capt-of}

\newcommand{\Be}{{Bragg-edge}}
\newcommand{\RaP}{{Ring-and-Plug}}
\newcommand{\rap}{{ring-and-plug}}
\newcommand{\CR}{{crushed ring}}
\newcommand{\lrt}{LRT}
\newcommand{\bepsilon}{\boldsymbol{\epsilon}}

\newcommand{\nhat}{{\hat{n}}}
\newcommand{\dd}{\; \textrm{d}}

\newcommand{\meas}{\Gamma_\epsilon}

\newcommand{\pder}[2]{\frac{\partial {#1}}{\partial {#2}}}
\newcommand{\ebasis}{\mathcal{E}}
\newcommand{\sbasis}{\mathcal{S}}
\newcommand{\pix}{p}

\begin{document}

\title{Tomographic Reconstruction of Two-Dimensional Residual Strain Fields from Bragg-Edge Neutron Imaging}

\author{A.W.T. Gregg}
\email[]{alexander.gregg@newcastle.edu.au}
\affiliation{School of Engineering, The University of Newcastle, Callaghan NSW 2308, Australia}

\author{J.N. Hendriks}
\affiliation{School of Engineering, The University of Newcastle, Callaghan NSW 2308, Australia}

\author{C.M. Wensrich}
\affiliation{School of Engineering, The University of Newcastle, Callaghan NSW 2308, Australia}

\author{A. Wills}
\affiliation{School of Engineering, The University of Newcastle, Callaghan NSW 2308, Australia}

\author{A.S. Tremsin}
\affiliation{Space Sciences Laboratory, University of California, Berkeley CA 94720, USA}

\author{V. Luzin}
\affiliation{ACNS, Australian Nuclear Science and Technology Organisation (ANSTO), Kirrawee NSW 2232, Australia}

\author{T. Shinohara}
\affiliation{Materials and Life Sciences Facility, Japan Proton Accelerator Research Complex, Tokai-mura, Ibaraki 319-1195, Japan}

\author{O. Kirstein}
\affiliation{School of Engineering, The University of Newcastle, Callaghan NSW 2308, Australia}
\affiliation{European Spallation Source, Lund 223 63, Sweden}

\author{M.H. Meylan}
\affiliation{School of Mathematical and Physical Sciences, The University of Newcastle, Callaghan NSW 2308, Australia}

\author{E.H. Kisi}
\affiliation{School of Engineering, The University of Newcastle, Callaghan NSW 2308, Australia}

\date{\today}

\begin{abstract}

Bragg-edge strain imaging from energy-resolved neutron transmission measurements poses an interesting tomography problem.  The solution to this problem will allow the reconstruction of detailed triaxial stress and strain distributions within polycrystalline solids from sets of Bragg-edge strain images. Work over the last decade has provided some solutions for a limited number of special cases.  In this paper, we provide a general approach to reconstruction of an arbitrary system based on a least squares process constrained by equilibrium.  This approach is developed in two-dimensions before being demonstrated experimentally on two samples using the RADEN instrument at the J-PARC spallation neutron source in Japan.  Validation of the resulting reconstructions is provided through a comparison to conventional constant wavelength strain measurements carried out on the KOWARI engineering diffractometer within ANSTO in Australia.  The paper concludes with a discussion on the range of problems to be addressed in a three-dimensional implementation.

\end{abstract}

\maketitle

\section{Introduction}

Energy-resolved neutron transmission techniques now provide a means for obtaining high-resolution images of strain within polycrystalline solids \cite{santisteban02b,tremsin12,tremsin11,WORACEK2018141}.  These techniques rely upon the relative shifts of abrupt changes in transmission rate as a function of wavelength -- known as {\Be}s -- the position of which are governed by diffraction.  

Detailed descriptions of this approach can be found elsewhere (e.g. \cite{santisteban02b,santisteban02}).  Briefly, the process involves the measurement of transmission spectra, typically using time-of-flight techniques at pulsed neutron sources (e.g. J-PARC in Japan, ISIS in the UK, or SNS in the USA).  Current detector technology is now able to perform such measurements simultaneously over arrays of individual pixels as small as 55 $\mu$m.  From this data, shifts in the position of observed {\Be}s relative to a reference stress-free sample provide a measure of strain.  

The salient points of such a measurement can be summarised as follows;

\begin{enumerate}
	\item{As with all diffraction-based techniques, strain measured in this way represents the elastic component alone.}
	\item{The measured strain is the normal component in the transmission direction of the neutron beam.}
	\item{Strain measured by each detector pixel represents a through-thickness average along the path of the corresponding ray.}
\end{enumerate}

The success of this approach and development of instruments and associated detector technologies has prompted activity focused on solving the associated tomographic reconstruction problem \cite{abbey09,abbey12,kirkwood15,lionheart15,wensrich16b,wensrich16a,hendriks2017,gregg2017}.  The aim is to provide a method analogous to conventional Computed Tomography by which the full triaxial strain distribution within a sample could be reconstructed from a sufficient set of \Be{} strain images. Note that this involves the reconstruction of a tensor field --- an inherently more complex task.  

Once developed, this approach has the potential to make a significant impact in a number of areas within experimental mechanics.  A prominent example concerns the assessment of residual stress fields in systems such as additively manufactured, laser clad, peened, welded, cast, forged and/or otherwise deformed components.  In each case, residual stress locked in by the manufacturing process has a critical impact on the strength and performance of the resulting parts.  Bragg-edge strain tomography promises a unique full-field approach to examining these systems over practical length scales.  

This task revolves around the inversion of the Longitudinal Ray Transform (\lrt) which represents an appropriate model of the measurement process \cite{lionheart15}.  While in general this is a three-dimensional problem, for simplicity we will consider only two dimensions in this paper.  

\begin{figure}[h!]
\begin{center}
    \includegraphics[width=0.65\linewidth]{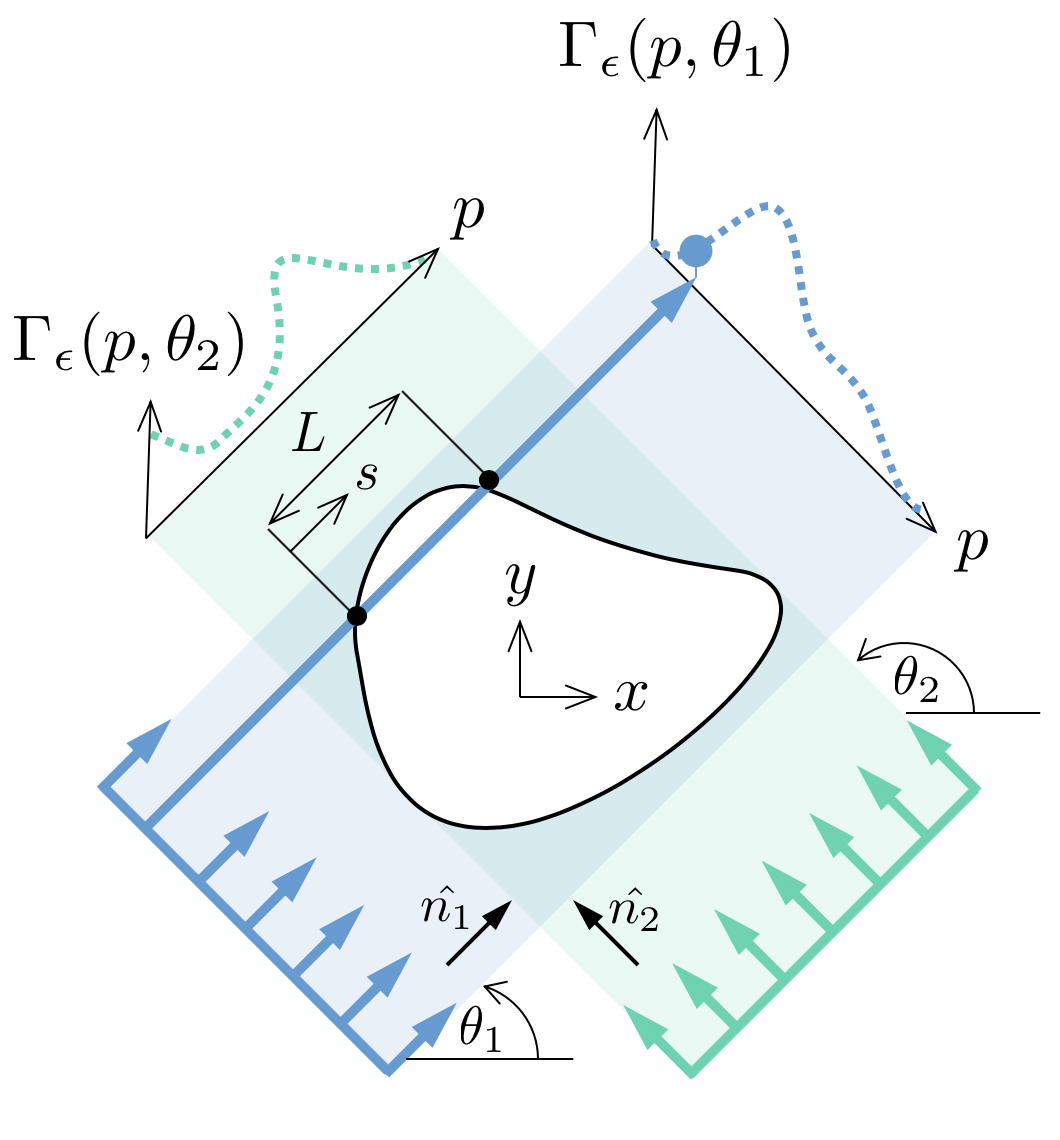}
    \caption{A single ray passes through a sample and provides a measurement of the through-thickness average normal strain in the direction of the ray at a detector pixel.  For each projection angle, $\theta$, measurements across the detector form a profile $\meas(\pix,\theta)$.}
    \label{fig:coord}
\end{center}
\end{figure}

With reference to the co-ordinate system and geometry shown in Figure \ref{fig:coord}, the \lrt{} can be written;
\begin{equation*}
\meas(\pix,\theta) =\frac{1}{L} \int _0^{L} \epsilon_{ij}(x(s,\pix),y(s,\pix)) \nhat_i \nhat_j \dd s,
\end{equation*}
where the rank-2 tensor strain field $\bepsilon$ is mapped to the average normal component of strain, $\meas$, along the ray with direction $\nhat = \begin{bmatrix}\cos({\theta}) & \sin({\theta}) \end{bmatrix}^T$ arriving at position $\pix$ on the detector.

From Lionheart and Withers \cite{lionheart15}, the LRT is known to be a non-injective mapping (from $\boldsymbol{\epsilon}(x,y)$ to $\meas(\pix,\theta)$).  Strain fields producing any given set of projections are not unique. As a consequence, general tomographic reconstruction is not possible from the measurements alone; additional information or constraints are required to isolate the correct (i.e. physical) field from all the possibilities.  To this end, a number of prior approaches have been developed that rely upon assumptions of compatibility or equilibrium to further constrain the problem.

Compatible strain fields are those that can be written as the gradient of a displacement field in a simply connected body (i.e. conservative).  In general, this is always the case.  However, when the total strain has both elastic and inelastic parts, only the compatibility of the sum is guaranteed.  If compatibility of the \emph{elastic} component can be assumed (e.g. in the absence of plasticity or other forms of eigenstrain), a strong constraint on the reconstruction problem exists. This constraint was central to the success of a number of prior reconstruction algorithms.

For example, the seminal work by Abbey \textit{et al.} \cite{abbey09, abbey12} on axisymmetric systems examined the reconstruction of strain within quenched cylinders and a standard VAMAS \rap{} sample using various basis functions alongside assumptions of compatibility.  Outside of axisymmetric systems, reconstructions have been demonstrated for a number of special cases; e.g. granular systems \cite{wensrich16b}, and strain fields resulting from \textit{in situ} loads \cite{wensrich16a,hendriks2017} where elastic strain compatibility can be assumed.

Unfortunately, in the vast majority of residual stress problems (e.g. all of the examples mentioned earlier), the elastic component of strain is inherently incompatible.  While compatibility cannot generally be assumed, equilibrium must always be satisfied.  Two separate algorithms for axisymmetric systems have been presented that rely on this assumption \cite{kirkwood15,gregg2017}.  

In the case of Kirkwood \textit{et al.} \cite{kirkwood15}, the assumption of equilibrium was not apparent at the time; it was a consequence of their approach to boundary conditions. In contrast, equilibrium was explicit and central to the method presented in \cite{gregg2017}.  Equilibrium is also central to the method presented in \cite{Jidling2018a} where the unknown strain is reconstructed using a machine learning technique known as a Gaussian process \cite{rasmussen2006}. This probabilistic method approaches the problem by considering strain as a distribution of Airy stress functions, which automatically satisfy equilibrium.

In this paper we develop an approach for reconstruction of arbitrary two-dimensional systems using an equilibrium constraint to provide unique solutions. The resulting algorithm is demonstrated in both simulation and on experimental data.  We also provide a brief discussion on the potential extension to three-dimensions.

\section{Approach}

The typical geometry for \Be{} strain imaging is shown in Figure \ref{fig:coord}.  In each orientation, $\theta_i$, a profile of the form $\meas(\pix,\theta_i)$ is measured across the width of the detector --- each detector pixel contributes one point to this profile. Inherent symmetry of the transform implies projections over 180 degrees are sufficient, however in practice measurements are usually taken over an entire revolution.  A complete set of profiles can be arranged to form a transformed image that resembles a traditional sinogram (e.g. Figure \ref{fig:sinogram_beam}). Given this \emph{strain-sinogram}, we seek to recover $\bepsilon$ from the infinite number of fields which potentially map to it.

Our approach is as follows:
\begin{enumerate}
\item Define a basis for the set of possible strain fields, $\ebasis$.  Elements of $\ebasis$ may not necessarily be physical (that is, they may not satisfy equilibrium). 
\item Compute the corresponding set of strain-sinograms, $\sbasis$, by mapping each element of $\ebasis$ through the \lrt. This forward projection involves numerical integration along ray paths.
\item  Through constrained least-squares fitting, find a linear combination from $\ebasis$ such that;
	\begin{enumerate}
		\item[--]{The corresponding combination from $\sbasis$ provides the measured strain-sinogram, and,}
		\item[--]{Equilibrium is satisfied at a sufficient number of test points.}
	\end{enumerate}
\end{enumerate}

In a numerical implementation, $\ebasis$ is composed of a finite number of elements.  Ideally this set should be orthogonal and ordered with increasing complexity to facilitate truncation. To this end, our approach employs a  two-dimensional Fourier basis to write each component of strain in the form; 
\begin{align*}
\epsilon_{ij}(x,y) = \sum_{a,b \in \mathbb{Z}} \alpha_{ij}^{a,b} \sin\bigg(&\frac{a\pi}{L}x\bigg)\sin\bigg(\frac{b\pi}{W}y\bigg) \\
+ \beta_{ij}^{a,b} &\sin\bigg(\frac{a\pi}{L}x\bigg)\cos\bigg(\frac{b\pi}{W}y\bigg) \\
+& \gamma_{ij}^{a,b} \cos\bigg(\frac{a\pi}{L}x\bigg)\sin\bigg(\frac{b\pi}{W}y\bigg) \\
&\quad + \eta_{ij}^{a,b} \cos\bigg(\frac{a\pi}{L}x\bigg)\cos\bigg(\frac{b\pi}{W}y\bigg)
\end{align*}
where $a$ and $b$ are wave numbers, $L$ and $W$ are characteristic dimensions of the geometry, and $\alpha_{ij}^{a,b} \ldots \eta_{ij}^{a,b}$ are unknown coefficients to be determined by the algorithm.

Truncating this basis to $n$ and $m$ wave numbers in the $x$ and $y$ directions respectively (i.e. $a \in [0,n]$, $b \in [0,m]$) gives $12nm+3$ tensor functions -- 4 sinusoids for each component of strain, 3 components for each permutation of wave numbers and 3 constant fields.  While the forward-mapping of these functions is potentially a large task, it can be done offline and ahead of time.  In other words, a library of basis pairs can be calculated prior to any experiment provided that the sample geometry is known.

Through Hooke's law, the equations of equilibrium can be written directly in terms of strain.  In two-dimensions, this relies upon either a plane-stress or plane-strain assumption.  For example, assuming plane-stress provides;
\begin{align*}
\pder{}{x} (\epsilon_{xx}+\nu \epsilon_{yy}) + \pder{}{y} (1-\nu)\epsilon_{xy} &= 0 \\
\pder{}{y} (\epsilon_{yy}+\nu \epsilon_{xx}) + \pder{}{x} (1-\nu)\epsilon_{xy} &=0
\end{align*}
where $\nu$ is Poisson's ratio.

Our algorithm imposes these two equations at a set of test points distributed over the interior of the sample.  At each point this provides a linear constraint on the unknown coefficients. 

The resulting constrained least-squares problem can be solved using a variety of techniques. Our algorithm utilises the \texttt{lsqlin} MATLAB intrinsic function.

Choice of $n$ and $m$ requires no a-priori knowledge of the system; the size of the basis can be chosen as the minimum required to capture the relevant features in the observed strain-sinogram. This can be assessed by examining the residual between the strain-sinogram and the fitted version; ideally no structure should be visible above random noise.  In a sense, in terms of the resulting reconstruction, $n$ and $m$ have some similarity to resolution, however they are certainly not the same.

\section{Demonstration --- Simulation}
\label{sec:sim}

\begin{figure}[!h]
\begin{center}
    \includegraphics[width=\linewidth]{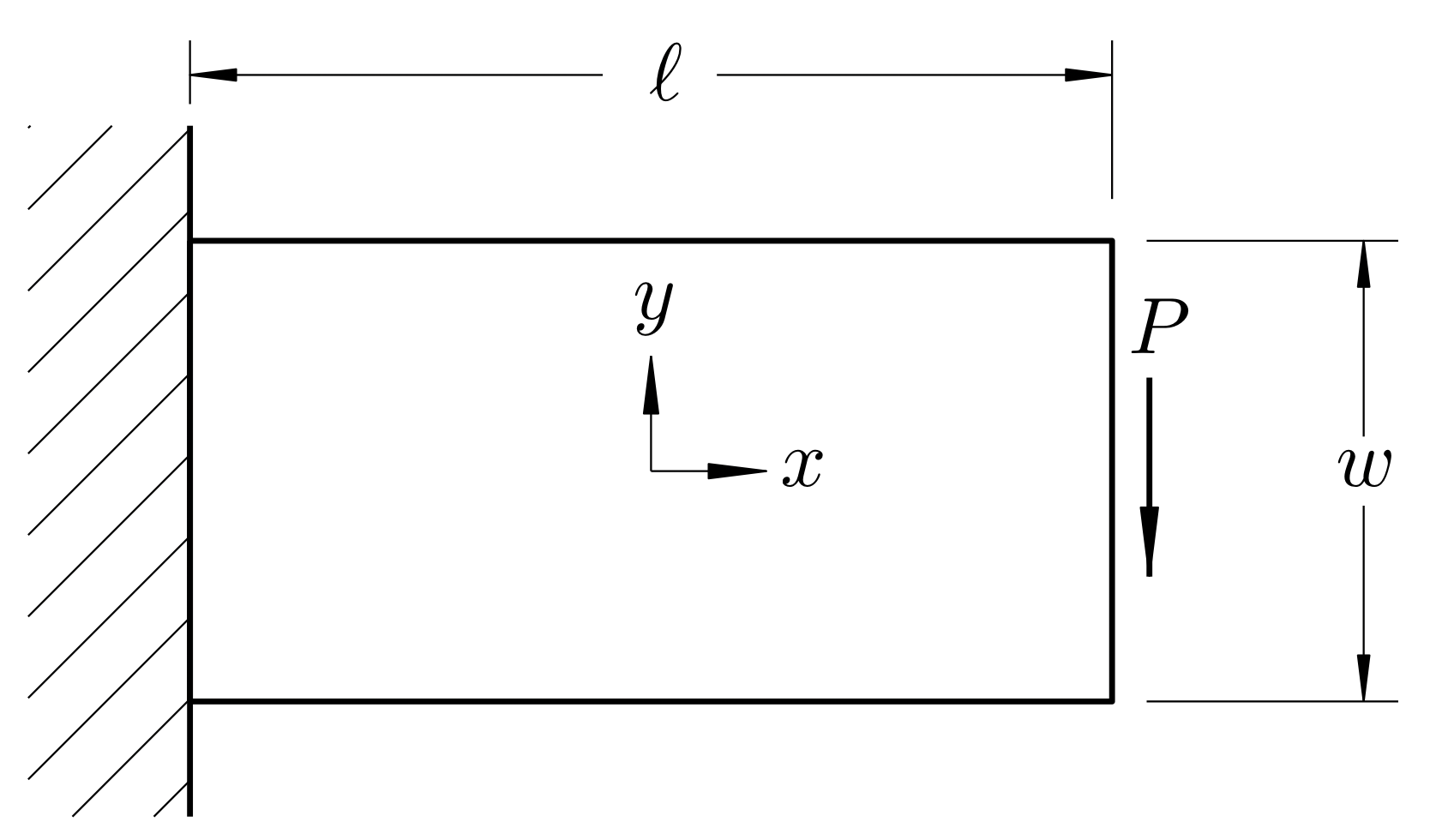}
    \caption{Cantilevered beam coordinate system and geometry. $\ell=20$ mm, $w=10$ mm, $P=2$ kN, $E=200$ GPa and $\nu=0.3$ \cite{wensrich16a}.}
    \label{fig:beam}
\end{center}
\end{figure}

We first demonstrate this algorithm on the classical cantilevered beam as examined previously by Wensrich et al. \cite{wensrich16a} and shown in Figure \ref{fig:beam}.  Under a plane-stress assumption, the Saint-Venant approximation to the resulting strain field is \cite{beer08}:
 \begin{equation*}
\bepsilon(x,y) =
\begin{bmatrix}
\frac{P}{EI}(\ell-x)y & -\frac{(1+\nu)P}{2EI}\left( \left(\frac{w}{2}\right)^2-y^2\right) \\
-\frac{(1+\nu)P}{2EI}\left( \left(\frac{w}{2}\right)^2-y^2\right) & -\frac{\nu P}{EI}(\ell-x)y
\end{bmatrix},
 \end{equation*}
where $I$ is the second moment of area, $P$ is the applied load, $E$ is Young's modulus and $\nu$ is Poisson's ratio. $\ell$ and $w$ are the dimensions shown in Figure \ref{fig:beam}.

Note that this strain field is compatible; a fact that was central to the previous approach. In contrast, no such assumption is made by the current algorithm.

50 \Be{} strain profiles over equally spaced angles between $0$ and $180^\circ$ were numerically simulated from this field assuming a state-of-the-art Micro-Channel Plate (MCP) detector with 512 pixels over 28 mm \cite{tremsin12}.  Gaussian measurement noise with standard deviation $\sigma=1.25\times10^{-4}$ was introduced; a value within the capabilities of current neutron instruments  \cite{hendriks2017}.

\begin{figure*}[h] 
\begin{center}
    \includegraphics[angle=0,width=0.9\linewidth]{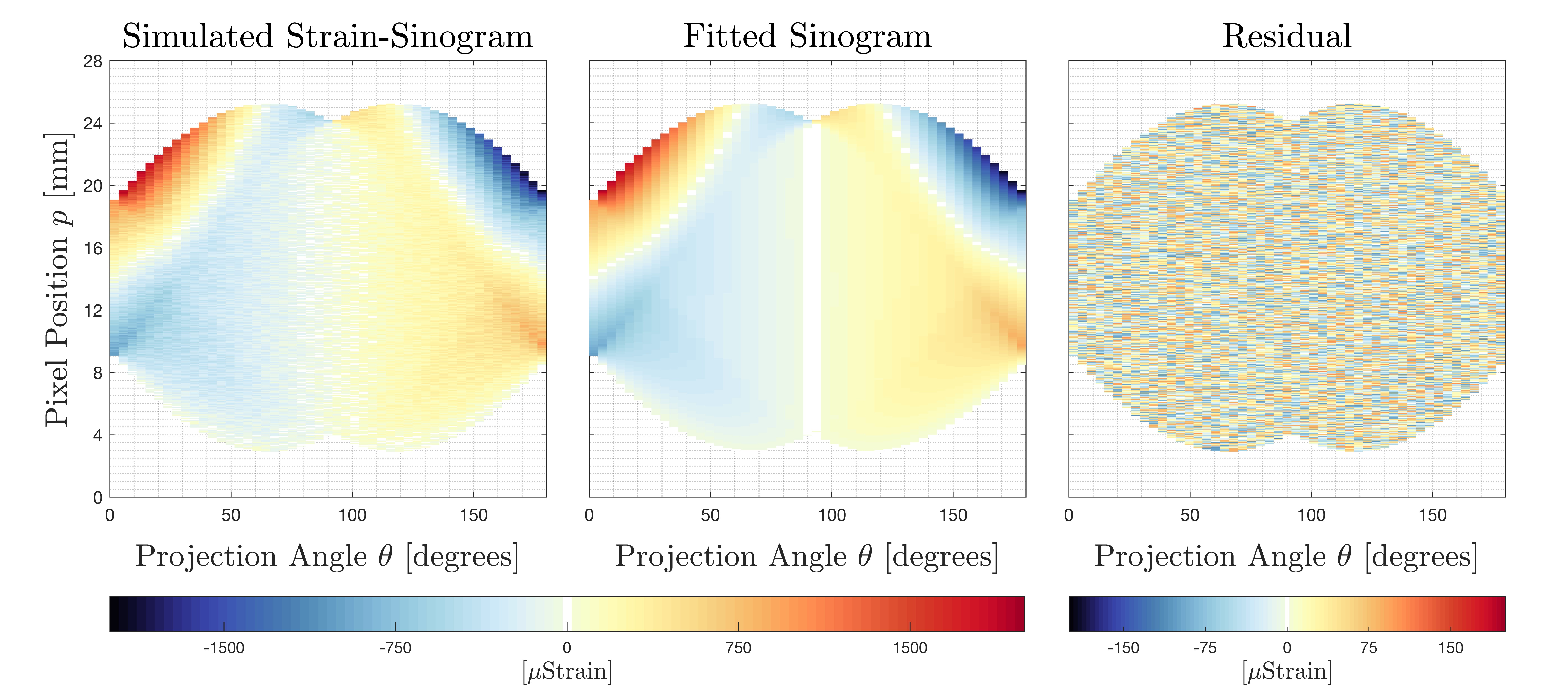}
    \captionof{figure}{(left) A simulated strain-sinogram from the cantilevered beam shown in Figure \ref{fig:beam}, (centre) the fitted strain-sinogram using 8 wave numbers in the $x$ and $y$ directions, and (right) spatial residual in the fit.}
    \label{fig:sinogram_beam}
\end{center}
\end{figure*}

The simulated strain-sinogram, resulting fit from $\sbasis$ and its residual based on $n=m=8$ wave numbers and a mesh of $1000$ equally spaced equilibrium test points is shown in Figure \ref{fig:sinogram_beam}. Characteristic lengths were chosen from the sample dimensions ($L=\ell$, $W=w$). It is clear that the residual has no structure, implying that a sufficient number of basis vectors have been used.

\begin{figure*}[h] 
\begin{center}
        \includegraphics[angle=0,width=0.9\linewidth]{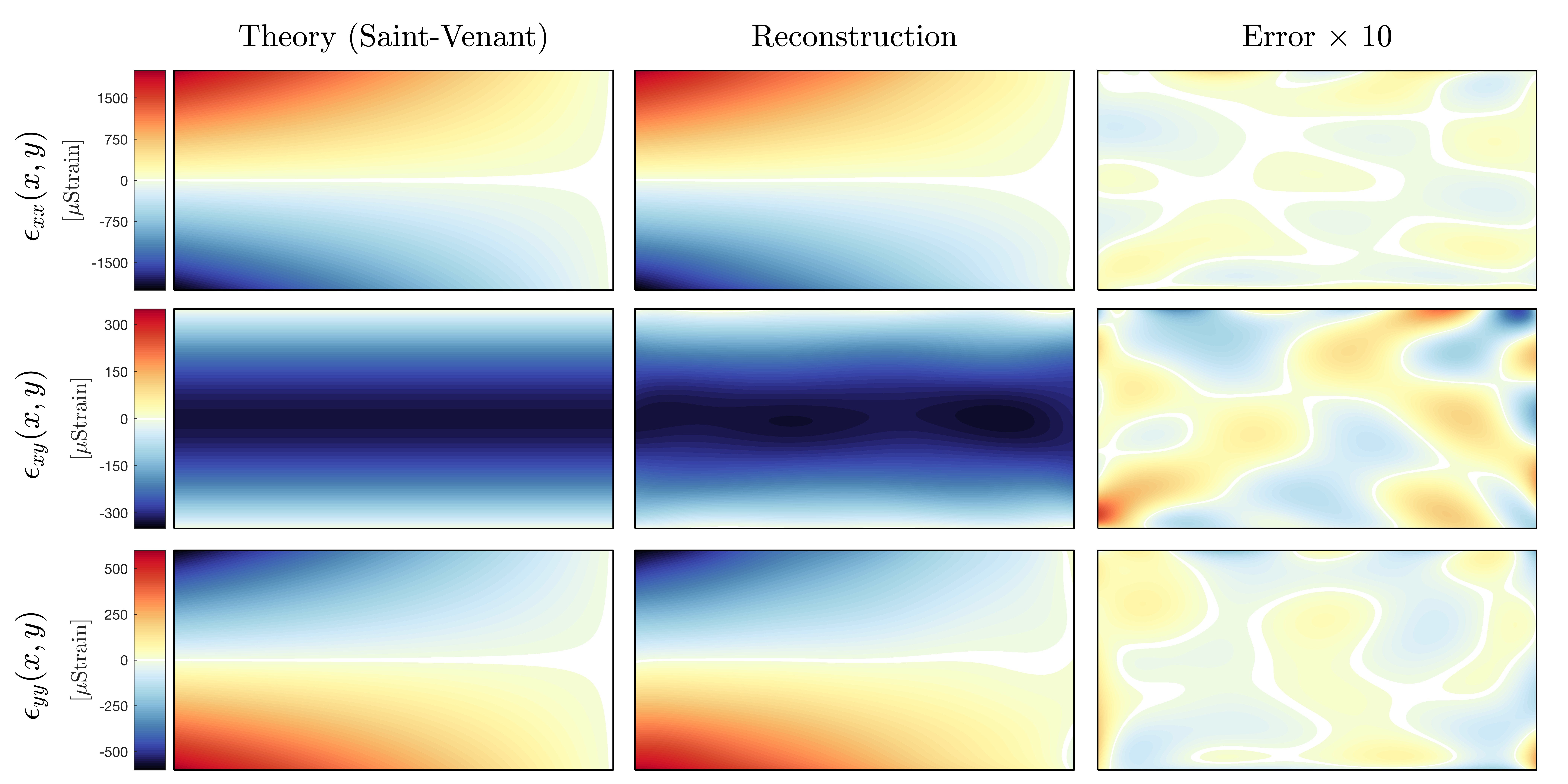}
    \captionof{figure}{(left) The Saint-Venant solution from which measurements were simulated, (centre) the reconstructed strain field, and (right) the error, scaled by a factor of 10. }
    \label{fig:recon_beam}
\end{center}
\end{figure*}

The resulting reconstruction in Figure \ref{fig:recon_beam} shows close agreement with the physical solution. Overall, the absolute error in strain is below $2.7\times10^{-5}$; almost one order of magnitude below the noise introduced into the measurements.  This would indicate that the mesh of equilibrium test points were sufficiently dense to isolate the physical solution.  Note that increasing the number of equilibrium points does not add significant computational burden, in fact in most cases the additional constraints aid the convergence.

\begin{figure}[!h]
\begin{center}
    \includegraphics[width=0.9\linewidth]{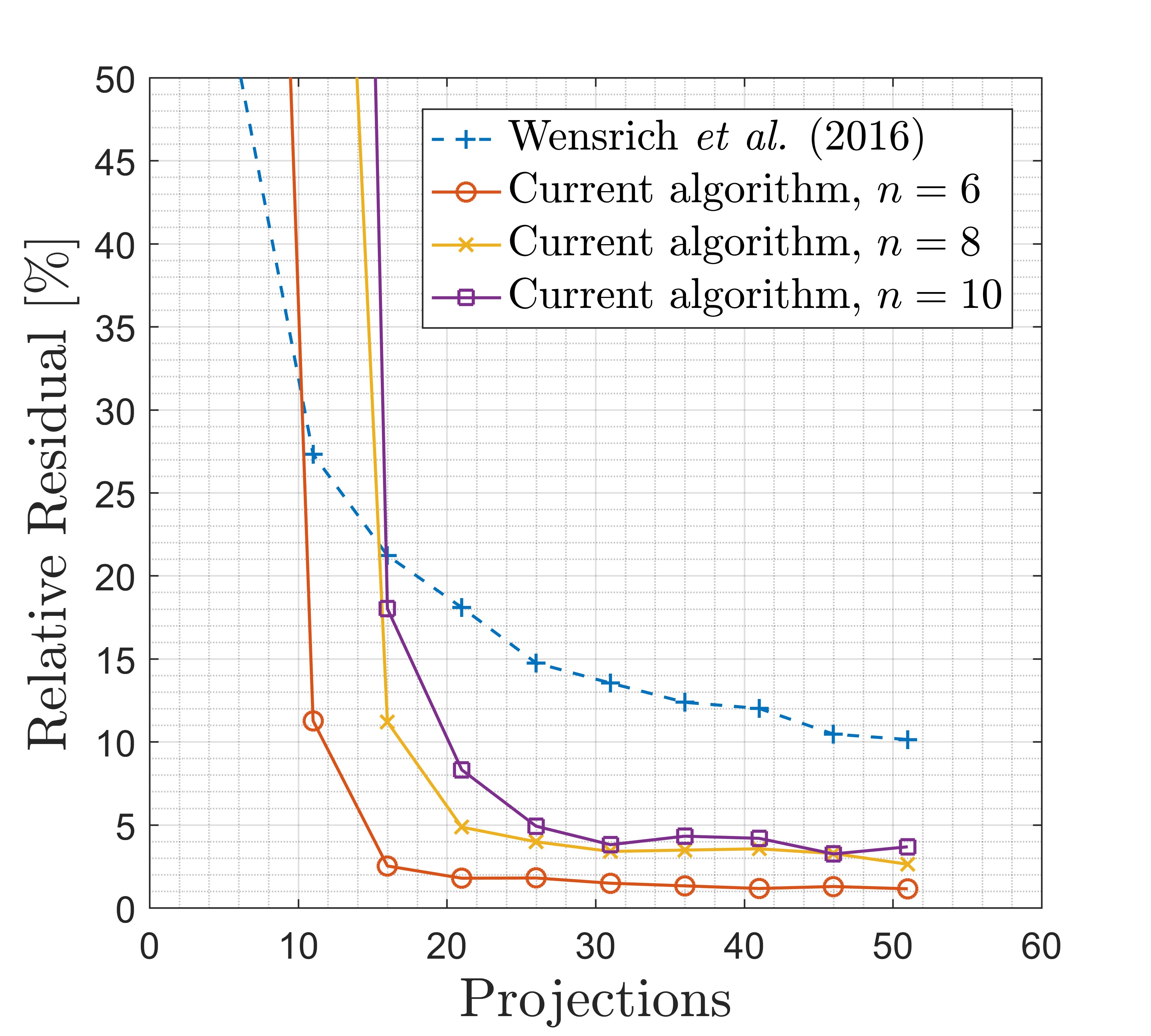}
    \caption{Convergence of the algorithm for the cantilevered beam as compared to the boundary reconstruction method presented in Wensrich \textit{et al.} \cite{wensrich16a}.}
    \label{fig:compari_plot}
\end{center}
\end{figure}

Direct comparison with the algorithm described by Wensrich \textit{et al.} \cite{wensrich16a} shows significantly faster convergence for this system (see Figure \ref{fig:compari_plot}). As expected, as the order of the basis increases the convergence is slower, however, even at $n=m=10$ the convergence is at least twice as fast.  Note that, with $n=m=10$, our problem involves 1203 unknown coefficients; far in excess of the 242 unknown boundary displacements in Wensrich \textit{et al.}.

\onecolumngrid

\twocolumngrid


\section{Demonstration --- Experimental}

Following success in simulation, the algorithm was demonstrated on real-world examples in an experiment on the RADEN energy resolved neutron imaging instrument at the Japan Proton Accelerator Research Complex (J-PARC) \cite{shinohara2016final,shinohara2015commissioning}.  This experiment focused on reconstructing residual strain fields within two EN26 steel samples (medium carbon, low-alloy) as follows;
\begin{enumerate}
	\item{A \CR{} formed through plastically deforming a hollow cylinder, and,}
	\item{An offset \rap{} system with residual strain resulting from an interference (i.e. shrink) fit.}
\end{enumerate}

\begin{figure}[!htb]
\begin{center}
    \includegraphics[width=\linewidth]{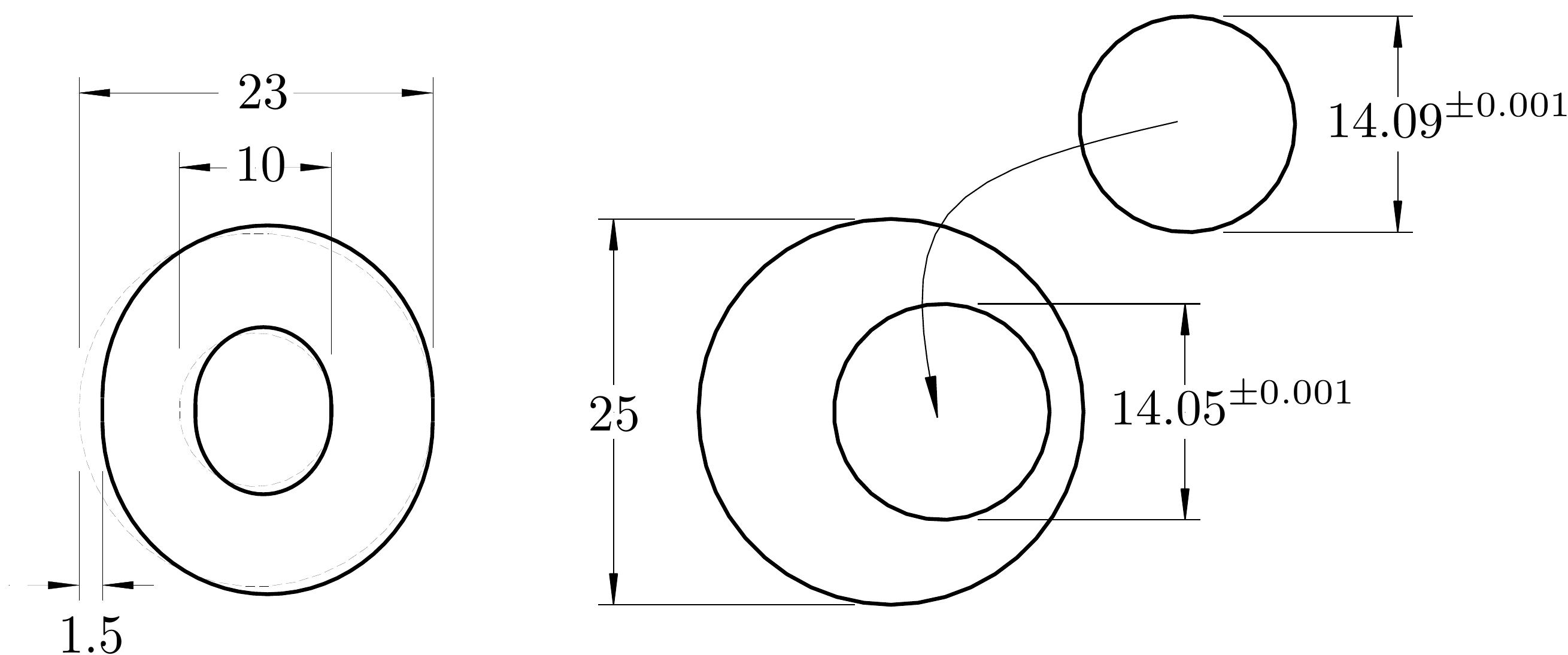}
    \caption{Sample geometries: (left) the \CR, and (right) the offset \rap.  All dimensions in mm.}
    \label{fig:crushedring}
\end{center}
\end{figure}

These samples were specifically designed to test the algorithm in the case of both continous (\CR) and discontinuous (ring-and-plug) strain fields.

Each sample was manufactured from the same bar of EN26 and was heat treated with an identical process to relieve stress and provide a uniform tempered-martensite structure (i.e. ferritic) prior to crushing/assembly. The final hardness of each sample was 290 HV.  Sample geometries are shown in Figure \ref{fig:crushedring}; both samples were 14 mm tall.  

The first sample was plastically deformed by $1.5$ mm on the diameter using approximately 8.4 kN of load from hardened steel platens in a mechanical testing machine.  

The second sample contained a total interference of $40\pm2$ $\mu$m produced through cylindrical grinding.  Finite element simulation suggested that this would provide strains of significant magnitude below yield.  After manufacture, the sample was assembled through a shrink-fit process (380$^\circ$ C versus -196$^\circ$ C).  

\begin{figure}[!htb]
\begin{center}
    \includegraphics[width=0.7\linewidth]{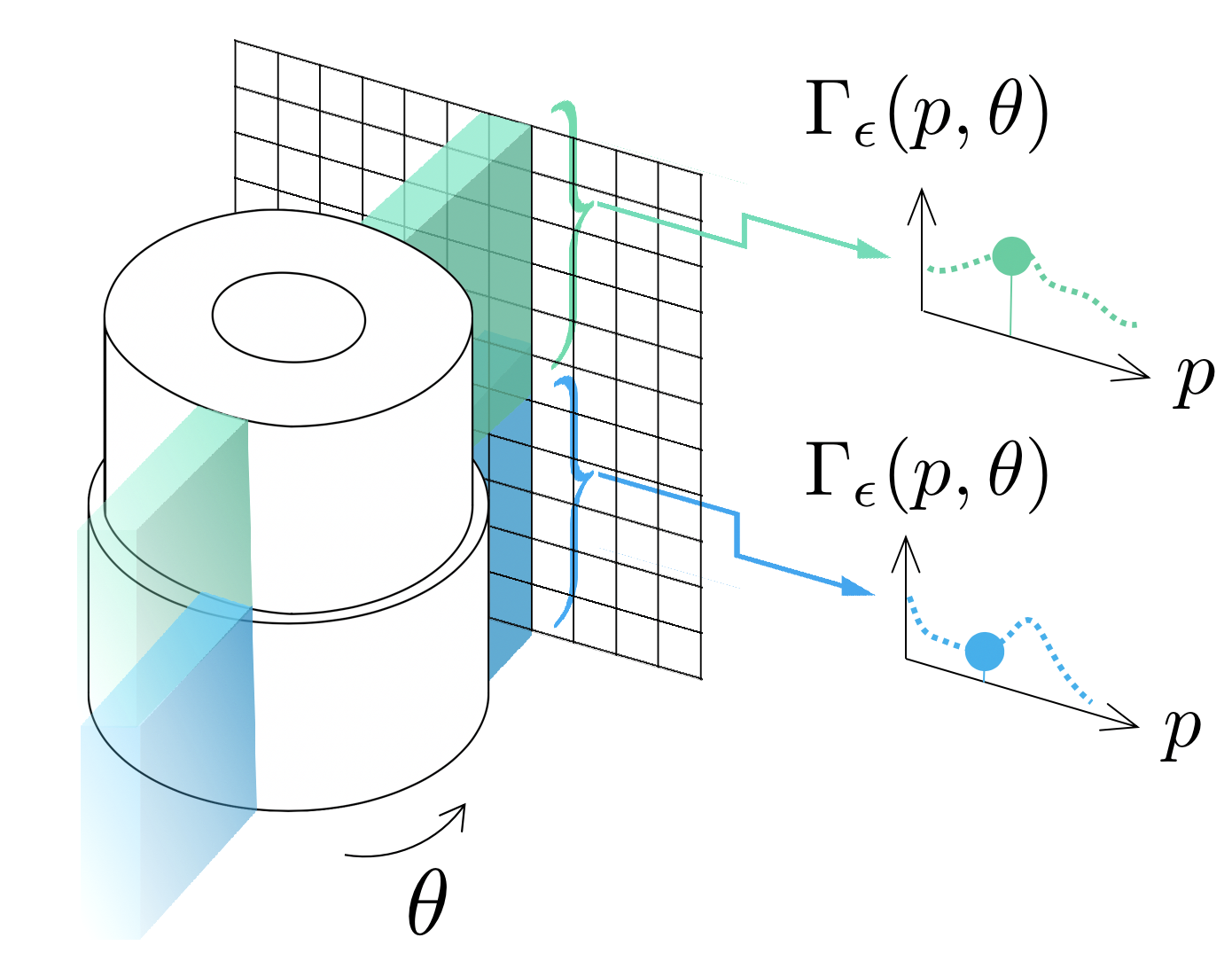}
    \caption{Neutron counts were binned over half-columns of pixels to provide a profile $\meas(\pix,\theta)$ from each sample.}
    \label{fig:stacked}
\end{center}
\end{figure}

Strain profiles were measured from both samples simultaneously using the RADEN instrument together with an MCP detector (512$\times$512 pixels, 55 $\mu$m per pixel) at a distance of 17.9 m from the source.  The source power was 409 kW (January 2018).  Counts were binned into  half-columns corresponding to the full height of each sample (one pixel wide) to provide the measured profiles $\meas(\pix,\theta)$ as shown in Figure \ref{fig:stacked}.   The resolution of the profiles was estimated from the sharpness of the sample boundaries and found to be approximately 100 $\mu$m. Note that this does not correspond to the resolution of the final reconstructions which, as mentioned earlier, is a more complicated matter.

\begin{figure}[!htb]
\begin{center}
    \includegraphics[width=0.9\linewidth]{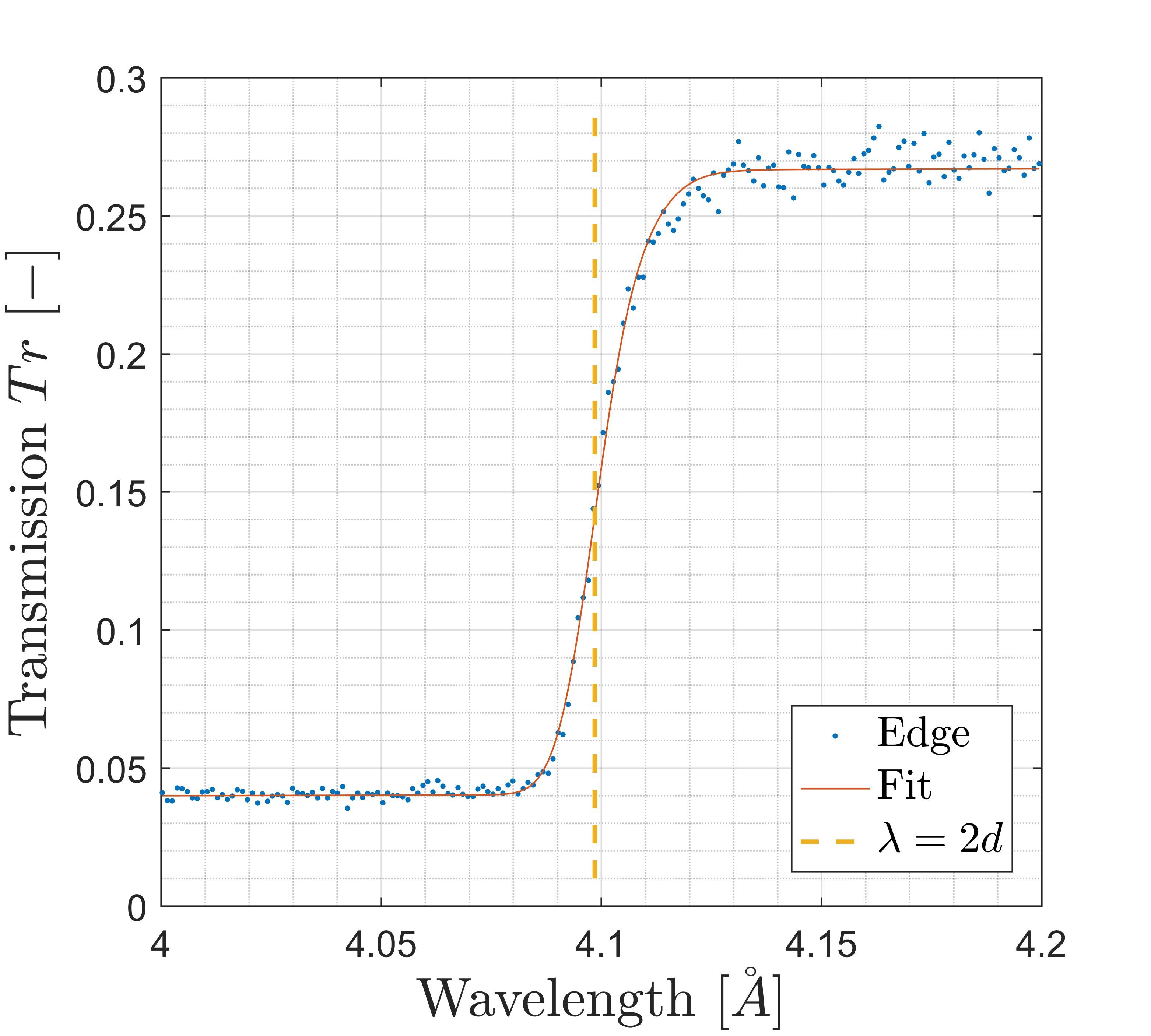}
    \caption{A typical measurement of the (110) Bragge-edge together with a fitted profile based on the Kropff model.}
    \label{fig:edgefit}
\end{center}
\end{figure}

Each individual strain measurement was of the form;
\begin{equation*}
	\bar{\epsilon} =\frac{d-d_0}{d_0},
\end{equation*}
where the atomic lattice spacing $d$ was found through fitting the integral form of the Kropff model to the (110) \Be, with $d_0$ the undeformed reference spacing (assumed constant). A typical edge fit is shown in Figure \ref{fig:edgefit}. A more detailed description of the fitting process is outlined in \cite{santisteban02b} and \cite{tremsin12}.

Throughout the experiment it was apparent that the fitted edge position was sensitive to sample thickness.  This effect has previously been described by Vogel \cite{vogel2000rietveld}, however the exact mechanism is yet to be established.  Potentially, the effect is a consequence of a weighting towards shorter wavelengths in the transmitted spectrum with sample thickness due to energy dependent attenuation -- generally known as beam hardening \cite{Brooks1976}. In our case, this may lead to a systematic bias in the observed location of edges depending on the path length. Along with a decrease in the height, beam hardening can slightly modify the shape of an edge and sensitivity between parameters in the curve fitting process can result in a perceived pseudostrain.
 
\begin{figure}[!htb]
\begin{center}
    \includegraphics[width=0.9\linewidth]{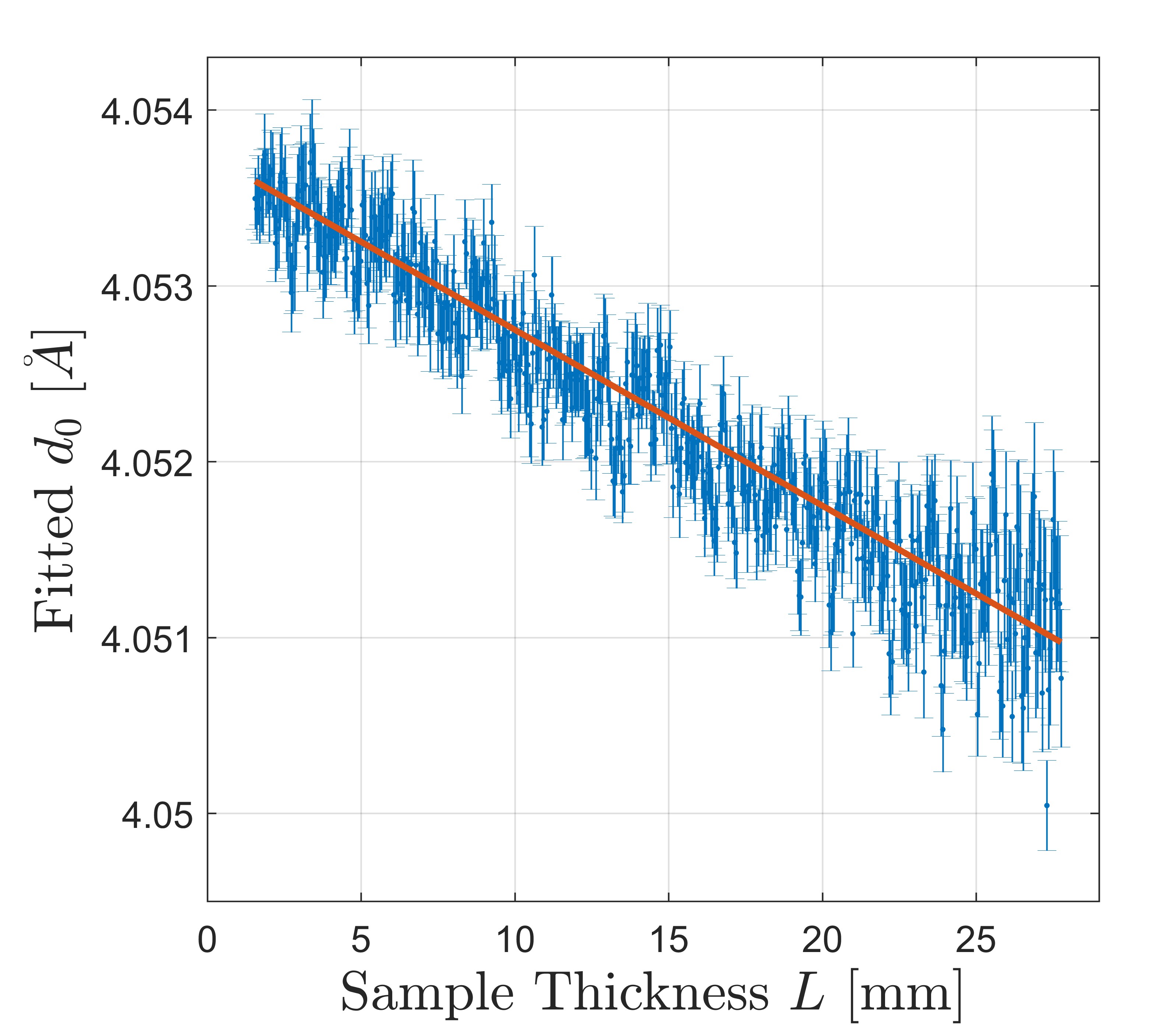}
    \caption{Bias in the fitted $d_0$ value as a function of the irradiated path length.}
    \label{fig:d0}
\end{center}
\end{figure}

To account for this effect, a correction was applied to $d_0$ as determined via a stress-relieved wedge-shaped sample.  \Be{} positions were measured from this sample over 9 hours allowing a linear trend against thickness to be determined as shown in Figure \ref{fig:d0}.

This empirical model proved sufficient for our purposes, however a more theoretical approach based on known neutron cross-sections is being investigated.  There is also the potential to approach this problem through full pattern fitting techniques as described by \cite{HUANG20072683,VOGEL20021,sato2011rietveld,sato2013upgrade,sato2017further}. Developments in this area of research have the potential to resolve many potential issues in the strain measurement process such as texture and grain size effects as well as this current issue.  At present this is not practical in terms of the number of individual measurements and the time required to fit a single pattern, however this will certainly improve in the future.

In total, 50 profiles were measured at golden angle increments \cite{feng2014golden} in $\theta$ with a sampling time of 2 hours per projection.   This provided a statistical uncertainty in strain of the order $1\times10^{-4}$ over most of the measurements. Together with open-beam and $d_0$ measurement, 4.5 days of beamtime were utilised.

Alignment of each sample was determined through matching the projected sample outlines to the conventional sinograms.  This involved calculating positions relative to the centre of rotation, and, in the case of the \CR , the initial angular offset.

Validation relied upon comparison to detailed conventional strain scans \cite{kisi2012applications,fitzpatrick2003analysis,noyan2013residual} from the KOWARI constant wavelength strain-diffractometer at the Australian Nuclear Science and Technology Organisation (ANSTO) \cite{kirstein2009strain,brule2006residual,kirstein2010kowari}.  These scans provided measurements of the three in-plane components of strain over a mesh of points within each sample (174 points in the crushed ring and 195 points in the offset ring-and-plug).  These were based upon the relative shift of the (211) diffraction peak measured using neutrons of wavelength $\lambda=1.67$ \AA{} ($90^\circ$ geometry) and a $0.5\times0.5\times14$ mm gauge volume. Note that the \{211\} and \{110\} lattice planes effectively have the same diffraction elastic constants \cite{DAYMOND20021613}. 

Sampling times on KOWARI were based on providing uncertainty in strain around $7\times10^{-5}$ which required around 30 hours of beamtime per component in the offset ring-and-plug and 15 hours per component in the crushed ring.  Together with sample setup and alignment, a total of 6 days of beamtime were required for the two samples.

\section{Results}

\begin{figure*}[!htb]
\begin{center}
    \includegraphics[angle=0,width=0.7\linewidth]{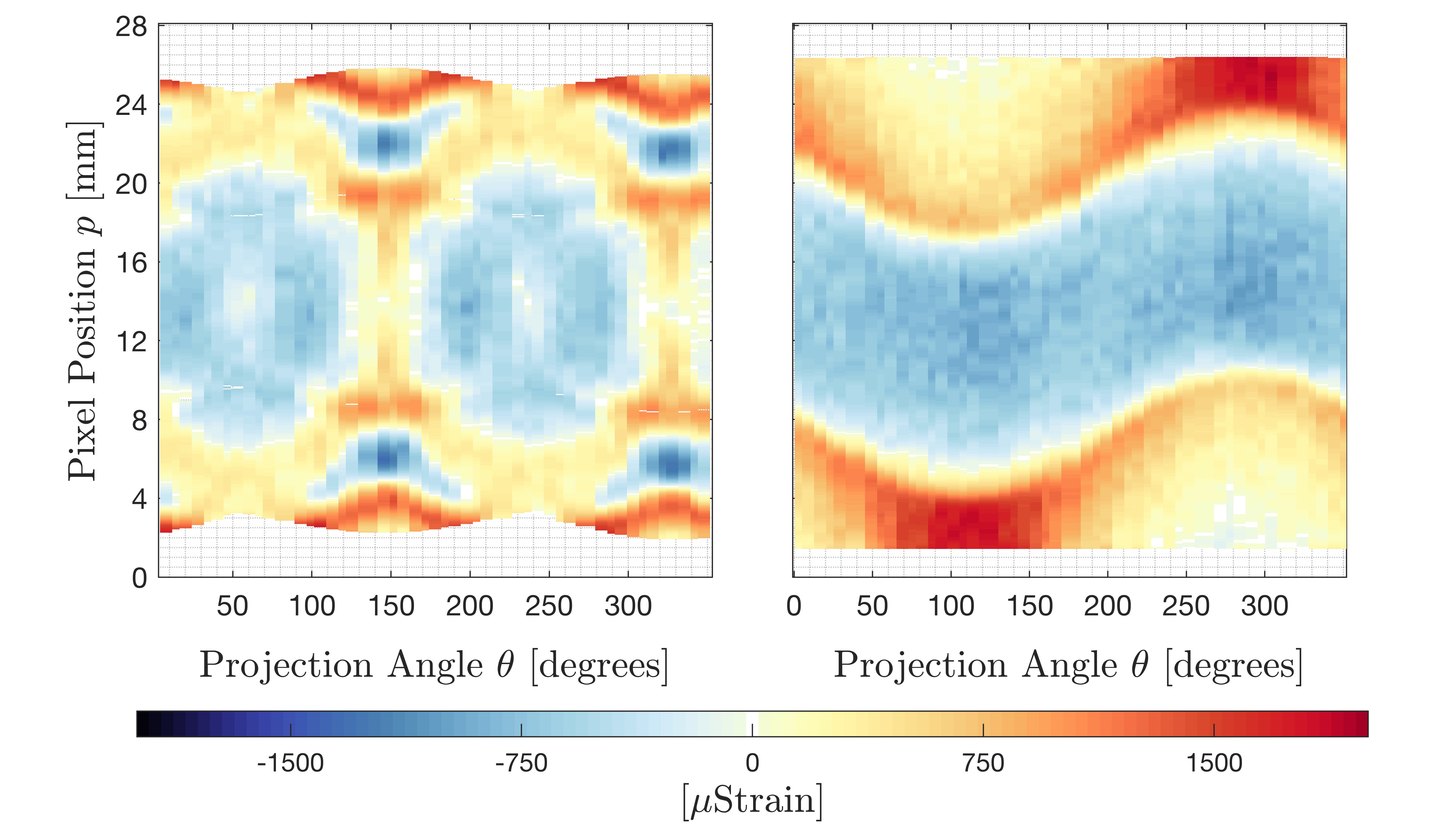}
    \caption{The measured strain-sinograms for the \CR{} (left) and \rap{} system (right).}
    \label{fig:sinogram_combined}
    \includegraphics[angle=0,width=0.9\linewidth]{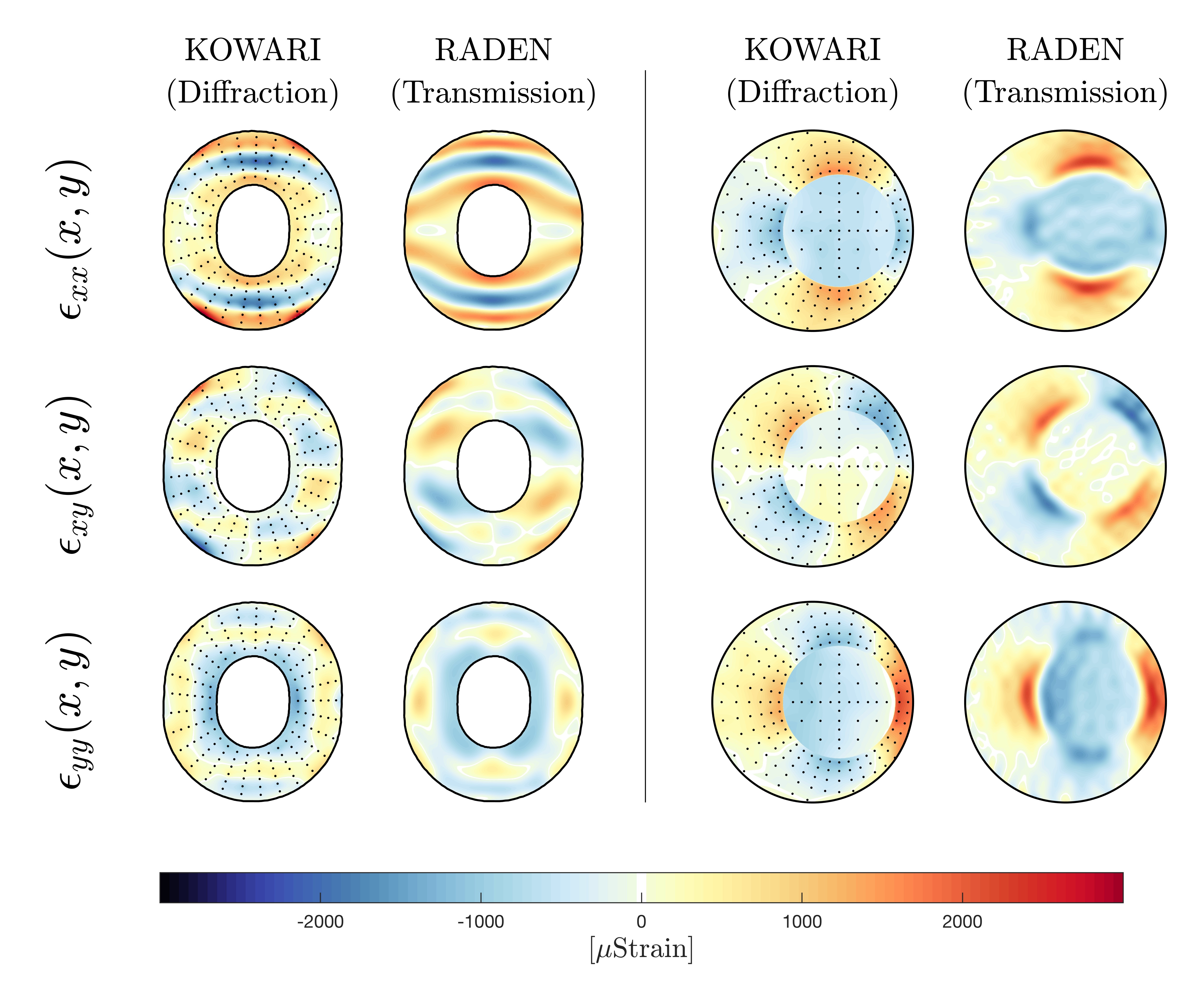}
    \caption{Strain maps interpolated from point-wise measurements on KOWARI compared to reconstructions from transmission measurements on RADEN for (left) the \CR{}, and (right) the ring-and-plug system.}
    \label{fig:recon_combined}
\end{center}
\end{figure*} 

\subsection{Crushed Ring}

\begin{figure}[htb!]
\begin{center}
    \includegraphics[width=\linewidth]{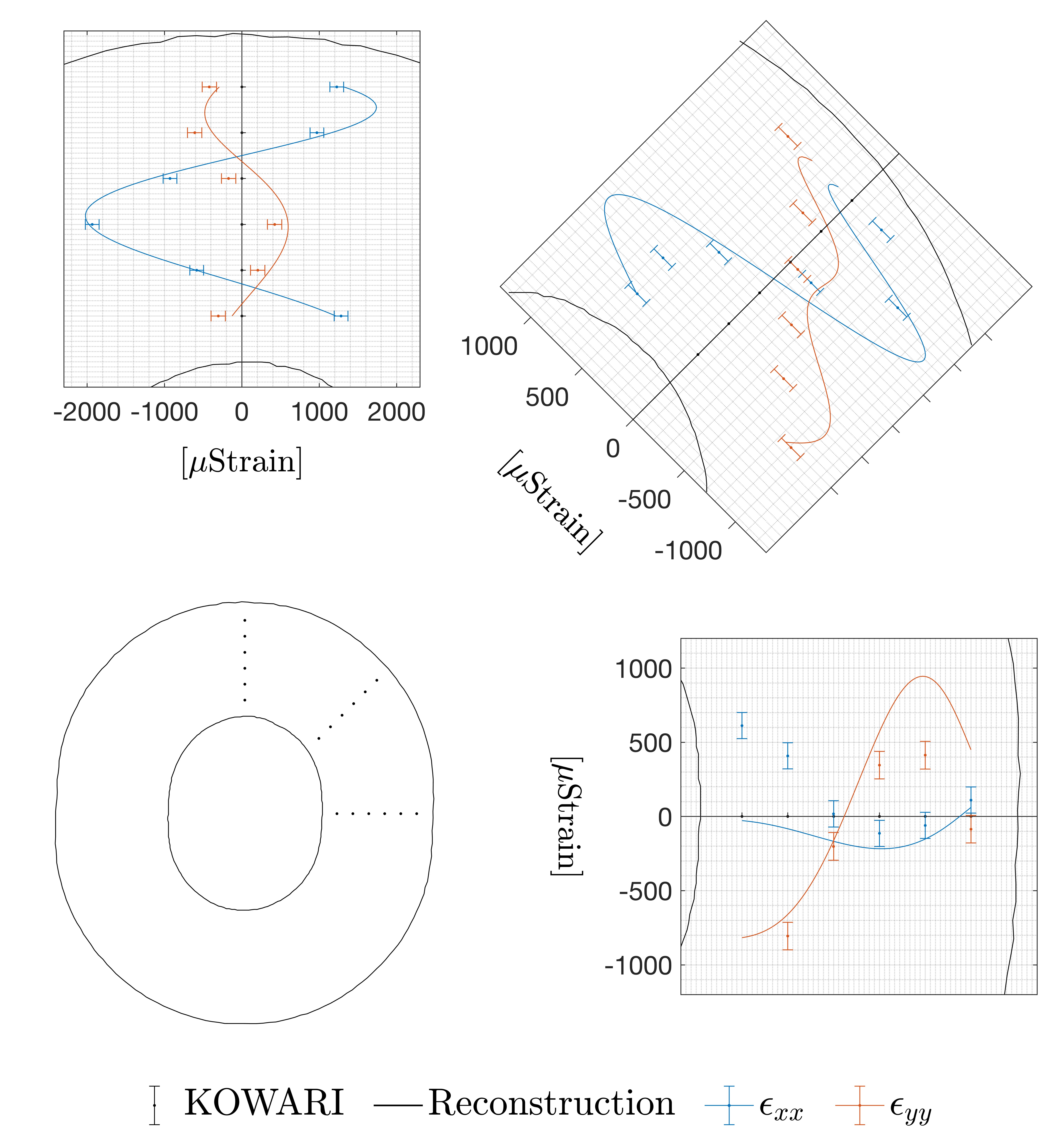}
    \caption{Distribution of $\epsilon_{xx}$ and $\epsilon_{yy}$ strain components over a number of cross sections within the crushed ring.}
    \label{fig:profiles_crushed}
\end{center}
\end{figure}

The measured strain-sinogram from the \CR{} is shown in the left-hand-side of Figure \ref{fig:sinogram_combined}.  Reconstruction from this data was carried out using 10 wave numbers in both the $x$ and $y$ directions and 1000 regularly spaced equilibrium test points over a grid on the interior of the sample. Characteristic lengths were chosen in-line with the major and minor axes of the crushed ring. The reconstructed strain field is shown on the left of Figure \ref{fig:recon_combined} compared to an interpolation of the KOWARI strain scans.  Figure \ref{fig:profiles_crushed} provides a direct comparison along a number of key cross sections.

In general, the reconstruction shows close agreement to the KOWARI measurement in terms of overall structure of the strain distribution. In particular, the symmetries present within the sample can be observed within the reconstruction despite the fact that no such assumption was made.  At a detailed level, there are some areas of discrepancy.  For example, the $\epsilon_{xx}$ component shows more pronounced banding across the width of the sample compared to KOWARI, and does not capture the full extent of the square-shaped tensile region in the $\epsilon_{yy}$ component.  

This behaviour was not observed in reconstructions based on simulated measurements from the interpolated KOWARI strain maps -- even with significant levels of simulated Gaussian noise. This suggests that the issue is not with the particular field or sample geometry, but systematic errors within the \Be{} fitting process. The validity of the plane-stress assumption (or lack thereof) may also play a role.

\subsection{Offset \RaP}

\begin{figure}[!htb]
\begin{center}
    \includegraphics[width=\linewidth]{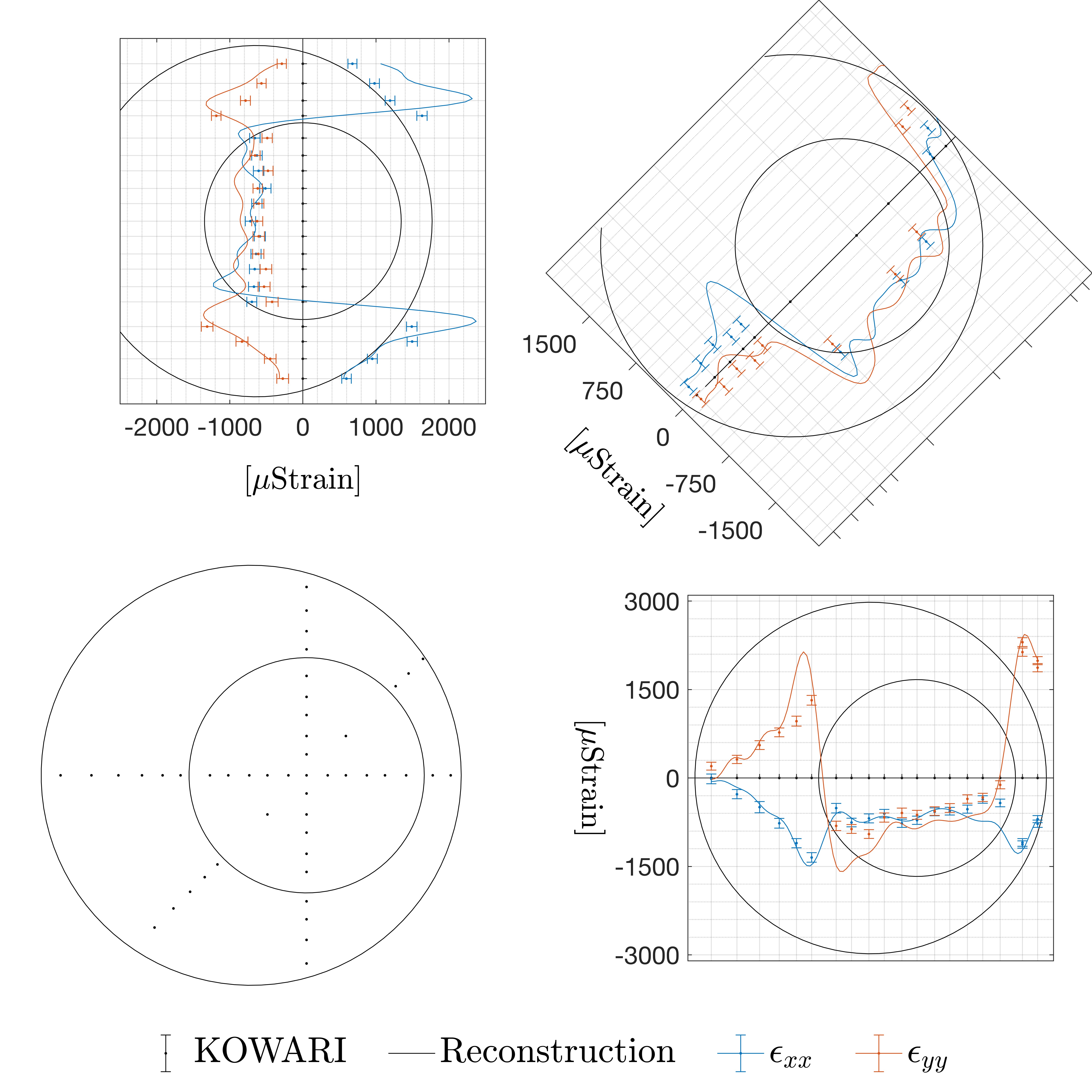}
    \caption{Distribution of $\epsilon_{xx}$ and $\epsilon_{yy}$ strain components over a number of cross sections within the ring and plug.}
    \label{fig:profiles_rp}
\end{center}
\end{figure}

The discontinuities in the \rap{} system necessitated the use of higher-order basis functions.  The reconstruction for this systems was based on 30 wave numbers in both the $x$ and $y$ directions (i.e. $n=m=30$) and characteristic lengths equal to the sample diameter. Equilibrium was enforced at 1000 equally spaced points. The right-hand-sides of Figures \ref{fig:sinogram_combined} and \ref{fig:recon_combined} show the measured strain-sinogram and reconstruction respectively. Figure \ref{fig:profiles_rp} shows a comparison over 3 key cross-sections.

As with the \CR , the reconstruction and KOWARI measurements show good overall agreement. The discontinuity in strain between the ring and plug obviously presents an interesting challenge with ringing artefacts clearly present in the reconstruction.  This effect is particularly evident in Figure \ref{fig:profiles_rp}, where overshoots and oscillations can be seen in the region of the step.  This effect was lessened by including higher order terms, however arbitrarily increasing $n$ and $m$ is not practical; the number of unknown coefficients grows with $12nm$ and can rapidly approach the number of measurements. Prior to this limit, the computational burden may become impractical. 

One potential solution is to use a `tailored' basis in which strains within the ring and plug are constructed from separate basis functions (e.g. \cite{gregg2017}). While this can eliminate the ringing, it is not a general approach since it requires prior knowledge about the composition of the system. In effect, the KOWARI measurements we are comparing to have been treated in this way; two separate interpolants have been used to generate the strain map shown in Figure \ref{fig:recon_combined}.  This is appropriate in this case, given that it serves as a reference with which to compare our reconstruction.  It should also be noted that this problem is a direct result of the discontinuity -- in the vast majority of practical cases strain fields tend to be smooth and this issue will not occur.

\subsection{Error Assessment}

From these results, a quantitative assessment of the discrepancy between the diffraction measurements and tomographic reconstructions was carried out.  In both cases, the difference was mean zero and Gaussian.  This would imply that the $d_0$ correction effectively removed the bias associated with sample thickness.

Over the 174 points measured within the crushed ring, the standard deviation of the difference was 370 $\mu$Strain.  Similarly, over the 195 points measured within the offset ring-and-plug, the standard deviation was 290 $\mu$Strain.  These are slightly higher than expectations based on the simulation results, however it should be pointed out that we are comparing to measurements which potentially have their own biases.

\section{Extension to Three Dimensions}
\label{sec:3D}

The algorithm outlined in this paper does not rely on the sample geometry being two-dimensional - in fact it can be easily extended to three dimensions with a small increase in complexity.

In three-dimensions there are six unknown components of strain to reconstruct. This obviously increases the computational burden associated with forward-mapping and fitting basis functions. For example, a real-valued three-dimensional Fourier series would entail $24n^3$ basis functions for $n$ wave numbers in each direction (as opposed to $12n^2$).  However, there is also an additional equation of equilibrium that provides a stronger constraint on any linear combination.  The amount of information per projection is also significantly increased; i.e. two-dimensional images versus one-dimensional profiles.  From this perspective, the number of projections required is likely to remain roughly equivalent for the same measurement resolution.  Note that, in three-dimensions, projections would need to be distributed over all directions in three-dimensional space.

Overall, the size of the problem would be larger, however the numerical approach would remain the same.

The true difficulty surrounds the implementation.  In three-dimensions, correspondingly larger sampling times are required to provide equivalent measurement uncertainty in two-dimensional images.  At present this would certainly require compromise in terms of the trade-off between measurement uncertainty and resolution through grouping multiple detector pixels.  

In the present work, columns of 256 pixels were grouped to provide one-dimensional profiles; to achieve the same uncertainty in a two-dimensional image, blocks of $16\times16$ detector pixels ($0.88\times0.88$ mm) would be required.  This situation may improve in the future as sources improve; e.g. J-PARC is expected to reach 800 kW in the near future with additional increases over 1 MW scheduled. Once commissioned, the European Spallation Source (ESS) in Sweden promises to be even brighter.  At 800 kW, image resolutions as low as $0.5 \times 0.5$ mm would be achievable with only a doubling of sampling time.  It should also be noted that, in the current work, we have erred on the side of caution in terms of the uncertainty-resolution compromise at the expense of sampling time; comparable results may have been possible with less beamtime.  

Given its importance, the effects of the uncertainty-resolution compromise forms a central question that must be investigated prior to three-dimensional implementation.

Associating each measurement with a defined path through known three-dimensional sample geometry also poses significant additional complexity.  This is coupled with the fact that more than one axis of rotation is required to view the sample from all directions with blind-spots potentially created by the positioning stage.  

If achieved, three-dimensional Bragg-edge tomography has the potential to provide information that cannot practically be measured any other way; full-field mapping in three-dimensions using current neutron strain scanners is a difficult process restricted by practical limitations in gauge volume size ($\approx$1 mm$^3$) and count times. 

In principle, the issues involved in three-dimensional strain tomography are not insurmountable and they form a natural focus for future work.

\section{Conclusion}

An algorithm for the reconstruction of biaxial elastic strain tensor fields from \Be{} neutron images has been presented. In contrast to previous algorithms, our method is capable of reconstructing residual strain since no assumption of elastic strain compatibility is made.

This approach was demonstrated in simulation and using experimental data collected from two samples on the RADEN energy-resolved neutron imaging instrument. Results showed excellent agreement with strain maps measured using the KOWARI constant wavelength engineering diffractometer.  

While Lionheart and Withers \cite{lionheart15} clearly demonstrated that Bragg-edge strain tomography is an ill-posed inverse problem, we have been able to achieve the task by considering the physical constraint imposed by equilibrium.  This experiment now represents the first ever tomographic reconstruction of residual strain fields outside of simple axisymmetric systems from \Be{} data.

At least in two-dimensions, full field Bragg-edge strain tomography can now provide a complementary approach to established pointwise diffraction-based strain measurement techniques.

The experiment has also highlighted a number of future areas of investigation.  These include the effects of beam hardening and strain gradients on the perceived elastic strain inferred from \Be s and the extension of the tomographic approach to three-dimensional strain fields.

\section{Acknowledgments}
This work is supported by the Australian Research Council through a Discovery Project Grant (DP170102324). Access to the RADEN and KOWARI instruments was made possible through the respective user access programs of J-PARC and ANSTO (J-PARC Long Term Proposal 2017L0101 and ANSTO Program Proposal PP6050).  The authors would also like to thank AINSE Limited for providing financial assistance (PGRA) and support to enable work on this project.

\bibliography{References}

\end{document}